\documentclass{article} 

\usepackage[paper=a4paper,left=3.0cm,right=3.0cm,top=4cm,bottom=4cm]{geometry}

\usepackage{amsmath,amssymb,amsthm}
\usepackage{epsfig,psfrag}
\usepackage{subfigure}
\usepackage{xspace}

\newcommand{\NN}{\ensuremath{\mathbb{N}}}

\newcommand{\etal}{{et al.}\xspace}
\newcommand{\wlg}{w.l.o.g.\xspace}

\newcommand{\calN}{\mathcal{N}}
\newcommand{\calR}{\mathcal{R}}  

\newtheorem{theorem}{Theorem}
\newtheorem{lemma}[theorem]{Lemma}
\newtheorem{corollary}[theorem]{Corollary}
\newtheorem{conjecture}[theorem]{Conjecture}
\newtheorem{definition}[theorem]{Definition}
\newtheorem{observation}[theorem]{Observation}

\begin{document}

\title{On the Convergence Time of the Best Response Dynamics in 
Player-specific Congestion Games\thanks{Parts of the results 
presented here already appeared in the Proceedings of the 4th Symposium on 
Stochastic Algorithms, Foundations, and Applications (SAGA) in 
2007~\cite{Ackermann:SAGA07}. }} \author{Heiner Ackermann \hspace{3ex} Heiko 
R\"oglin \\ Department of Computer Science \\ RWTH Aachen, D-52056 Aachen, 
Germany \\
$\{$ackermann, roeglin$\}$@cs.rwth-aachen.de }
\date{\today}

\maketitle

  \thispagestyle{empty}

  \begin{abstract}
We study the convergence time of the best response dynamics in player-specific 
singleton congestion games. It is well known that this dynamics can cycle, 
although from every state a short sequence of best responses to a Nash 
equilibrium exists. Thus, the random best response dynamics, which selects the 
next player to play a best response uniformly at random, terminates in a Nash 
equilibrium with probability one. In this paper, we are interested in the 
expected number of best responses until the random best response dynamics 
terminates.

As a first step towards this goal, we consider games in which each player can 
choose between only two resources. These games have a natural representation as 
(multi-)graphs by identifying nodes with resources and edges with
players. For the class of games that can be represented as trees, we show that the 
best-response dynamics cannot cycle and that it terminates after $O(n^2)$ steps 
where $n$ denotes the number of resources. For the class of games represented 
as cycles, we show that the best response dynamics can cycle. However, we also 
show that the random best response dynamics terminates after $O(n^2)$ steps in 
expectation.

Additionally, we conjecture that in general player-specific singleton
congestion games there exists no polynomial upper bound on the expected number 
of steps until the random best response dynamics terminates. We support our 
conjecture by presenting a family of games for which simulations indicate 
a super-polynomial convergence time.
\end{abstract}
  \newpage

\section{Introduction}

We study the convergence time of the best response dynamics to pure Nash
equilibria\footnote{In the following, the term \emph{Nash equilibrium}
always refers to a pure one.} in player-specific singleton congestion
games. In such games, we are given a set of resources and a set of players. Each
player is equipped with a set of non-decreasing, player-specific delay
functions which measure the delay the player experiences from allocating
a particular resource and sharing it with a certain number of other
players. A player's goal is to allocate a \emph{single} resource with
minimum delay given fixed choices of the other players.
Milchtaich~\cite{Milchtaich:JGEB96}, who was the first to consider
player-specific singleton congestion games, proves that every such game
possesses a Nash equilibrium which can be computed efficiently. However,
he also observes that these games are no potential
games~\cite{MondererShaply:JGEB96}, that is, the best response dynamics,
in which players consecutively change to resources with minimum delay,
can cycle. This is in contrast to congestion games with common delay
functions in which all players sharing a resource observe the same delay.
In the following, we refer to congestion games with common delay
functions as \emph{standard congestion games}.
Rosenthal~\cite{Rosenthal:JGame73}, who introduces standard congestion
games, proves that they always admit a potential function guaranteeing
the existence of Nash equilibria and that the best response dynamics
cannot cycle. Ieong \etal~\cite{Ieong_etal:AAAI05} consider the
convergence time of the best response dynamics to Nash equilibria in
standard singleton congestion games. They observe that the delay values
can be replaced by their ranks in the sorted list of theses values
without affecting the best responses dynamics. By applying Rosenthal's
potential functions to these new delay functions they observe that after
at most $n^2m$ best responses a Nash equilibrium is reached, where $n$
equals the number of players and $m$ the number of resources. This result
is independent of any assumption on the ordering according to which
players change their strategies.

Since the best response dynamics in player-specific singleton congestion games 
can cycle, we propose to study random best response dynamics in such games. 
This approach is motivated by the following observation due to 
Milchtaich~\cite{Milchtaich:JGEB96}: From every state of a
player-specific singleton congestion game there exists a polynomially
long sequence of best responses leading to a Nash equilibrium. Thus, the random best response 
dynamics selecting the next player to play a best response at random
terminates with probability one after a finite number of steps. 
Milchtaich's analysis leaves open the question how long it takes until the 
random best response dynamics terminates in expectation. In this paper, we 
address this question as we think that it is a natural and interesting one. 
Currently, we are not able to analyze the convergence time in arbitrary 
player-specific singleton congestion games. However, our
experimental results support the following conjecture.

\begin{conjecture}\label{Conjecture:PS_lowerBound}
  There exist player-specific singleton congestion games in which
  the expected number of steps until the random best response dynamics
  terminates is super-polynomial.
\end{conjecture}

In order to gain insights into the random best response dynamics, we
begin with very simple yet interesting classes of games, and consider
games in which each player chooses between only two alternatives. These
games can be represented as multi-graphs: each resource corresponds to a
node and each player to an edge. In the following, we call games that can
be represented as graphs with topology $t$ player-specific congestion
games on topology $t$. First, we consider games on trees and circles.

We prove that player-specific congestion games on trees admit a potential 
function from which we derive an upper bound of $O(n^2)$ on the maximum number 
of best responses until a Nash equilibrium is reached. Note, that this 
result is independent of the initial state and any assumption on the ordering 
according to which players change their strategies.

The result bases on the observation that one can replace the
player-specific delay functions by common delay functions without
changing the players preferences. Thus, player-specific congestion games
on trees are isomorphic to standard congestion games on trees and we can
apply the result of Ieong \etal~\cite{Ieong_etal:AAAI05} to upper bound
the convergence time. We proceed with player-specific congestion games on
circles, and show that these games are the simplest games in which the
best response dynamics can cycle. As we are only given four different
delay values per player, we characterize with respect to the ordering of
these four values in which cases the best response dynamics can cycle.
We observe that only one such case exists. Finally, we analyze the
convergence time of the random best response dynamics in such games, and
prove a bound of $O(n^2)$ on the expected number of steps until the
dynamics terminates. In order to prove this result we introduce the
notion of over- and overload tokens. An overload token indicates that a
resource is shared by two players, an underload token indicates that it
is unused. We observe that the number of tokens cannot increase, and that
once in a while tokens get stuck or vanish.

Based on the insights gained by analyzing player-specific congestion games on 
circles we present a family of games and conjecture that there exists no 
polynomial upper bound on the expected time until the random best response 
dynamics terminates. Obviously, this depends on the initial state, and so we 
implicitly assume that the initial configuration is chosen appropriately. Our 
conjecture is motivated by a slightly different notion of over- and underload 
tokens. Now their definition depends on the fact that every resource
has a fixed congestion that it takes in every Nash equilibrium.
In contrast to games on 
circles we show that the number of over- and underload tokens can also increase 
if the initial configuration is chosen appropriately. Intuitively one may think 
of the number of tokens as a measure of derangement of order. In games on 
circles this measure can only decrease whereas it can also increase in
general games. We fail to give a rigorous proof of a super-polynomial 
lower bound. However, we support our conjecture by empirical results obtained 
from simulations.

\subsection{Definitions and Notations}

A \emph{player-specific singleton congestion game} $\Gamma$ is a tuple
$(\calN, \calR, (\Sigma_i)_{i \in \calN}, (d_r^i)_{r \in \calR}^{i \in \calN})$ where 
$\calN $ denotes the set of $n$ players, $\calR$ the set of $m$ resources, 
$\Sigma_i \subseteq \calR$ the strategy space of player $i$, and $d_r^i \colon 
\NN \rightarrow \NN$ a strictly increasing  delay function associated with 
player $i$ and resource $r$. In the following, we assume that ties 
are broken arbitrarily. That is, for 
every pair of resources $r_1, r_2 \in \Sigma_i$ and every pair $n_{r_1}, 
n_{r_2} \in \NN$, $d_{r_1}^i(n_{r_1}) \neq d_{r_2}^i(n_{r_2})$. We
denote by $S=(r_1, \ldots, r_n)$ the \emph{state of the game} in which player $i$ 
allocates resource $r_i \in \Sigma_i$. For a state $S$, we define the 
\emph{congestion} $n_r(S)$ on resource $r$ by $n_r(S) = |\{i \mid r = r_i\}|$, 
that is, $n_r(S)$ equals the number of players sharing resource $r$ in state 
$S$. We assume that players act selfishly and wish to 
allocate resources minimizing their individual delays. The delay of player $i$ from allocating 
resource $r$ in state $S$ is given by $d_r^i(n_r(S))$. Given a state $S=(r_1, 
\ldots, r_n)$, we call a resource $r^* \in \Sigma_i \setminus \{r_i\}$ a 
\emph{best response} of player $i$ to $S$ if, for all $r' \in \Sigma_i 
\setminus \{r_i\}$, $d_{r^*}^i(n_{r^*}(S)+1) \le d_{r'}^i(n_{r'}(S)+1)$, and if 
$d_{r^*}^i(n_{r^*}(S)+1) \le d_{r_i}^i(n_{r_i}(S))$. Furthermore, we call $r_i$ 
a \emph{best response} of player $i$ to $S$ if, for all $r' \in \Sigma_i 
\setminus \{r_i\}$, $d_{r_i}^i(n_{r_i}(S)) \le d_{r'}^i(n_{r'}(S)+1)$. The 
standard solution concept in player-specific singleton congestion games
are \emph{Nash equilibria}. A state $S$ is a Nash equilibrium if for each player 
$i$ the resource $r_i$ is a best response.

In this paper, we consider games that have natural representations as graphs. 
We assume that each player chooses between only two resources.  In this case, 
we can represent the resources as the nodes of a graph and the players as the 
edges. If different players choose between the same two resources, then the 
corresponding graph has multi-edges. The direction of an edge naturally 
corresponds to the strategy the player currently plays.

In the following, we will sometime refer to \emph{standard singleton 
congestion games}. Standard singleton congestion games are defined in the same 
way as player-specific singleton congestion games except that we are not given 
player-specific delay functions $d_r^i$, $r\in \calR, i \in \calN$, but common 
delay functions $d_r$, $r\in \calR$. Ieong \etal~\cite{Ieong_etal:AAAI05}  
observe that in standard singleton congestion games one can always 
replace the delay values $d_r(n_r)$ with $r \in \calR$ and $1 \le n_r \le n$ by 
their ranks in the sorted list of these values without affecting the players 
preferences in any state of the game. Note that this approach is not restricted 
to standard singleton congestion games but also applies to player-specific 
singleton congestion games. That is, given a player-specific congestion game 
$\Gamma$, fix a player $i$ and consider a list of all delays $d_r^i(n_r)$ with 
$r \in \calR$ and $1 \le n_r \le n$. Assume that this list is sorted in a 
non-decreasing order. For each resource $r$, we define an alternative 
player-specific delay function $\tilde{d}_r^i:\NN \rightarrow \NN$ where, for 
each possible congestion $n_r$, $\tilde{d}_r^i(n_r)$ equals the rank of the 
delay $d_r^i(n_r)$ in the aforementioned list of all delays. Due to our 
assumptions on the delay functions, all ranks are unique. In the following, we 
define the \emph{type of a player $i$} by the ordering of the player-specific 
delays $d_{r}^i(1), \ldots,$  $d_{r}^i(n)$ of the resources $r \in \Sigma_i$.

We define the \emph{transition graph} $TG(\Gamma)$ of a player-specific 
singleton congestion game $\Gamma$ as the graph that contains a vertex for 
every state of the game. Moreover, there is a directed edge labeled with $i$ 
from state $S$ to state $S'$ if we obtain $S'$ from $S$ by permitting player 
$i$ to play a best response in $S$.

We call the dynamics in which players iteratively play best responses the 
\emph{best response dynamics}. Furthermore, we use the term \emph{best response 
schedule} to denote an algorithm that selects, given a state $S$, the next 
player to play a best response. We assume that such a player is always selected 
among those players who have an incentive to change their strategy. The 
convergence time $t(n,m)$ of a best response schedule is the maximum number of 
steps to reach a Nash equilibrium in any game with $n$ players and $m$ 
resources and for any initial state. If the schedule selects the next player to 
play a best response uniformly at random then $t(n,m)$ refers to the maximum 
expected convergence time. We use the term \emph{random best response
dynamics} to denote the resulting dynamics. Additionally, we use the
terms best response dynamics and best response schedule interchangeably.

\subsection{Related Work}

We already mentioned that every player-specific singleton 
congestion game possesses a Nash equilibrium which can be computed efficiently. 
Moreover, we mentioned that such games are no potential game, even
though from every state there exists a polynomially long sequence of
best responses leading to a Nash equilibrium. These results are due to 
Milchtaich~\cite{Milchtaich:JGEB96}. Milchtaich also observes that 
player-specific network congestion games, i.e., games in which each player 
wants to allocate a path in a network, do not possess Nash equilibria in 
general~\cite{Milchtaich:WINE06}. He proposes to characterize those games with 
respect to their networks which always possess Nash equilibria. Such a 
characterization should be independent of further assumptions on the delay 
functions. Ackermann, R\"oglin, and V\"ocking~\cite{Ackermann_etal:WINE06} 
extend the results presented in~\cite{Milchtaich:JGEB96} to player-specific 
matroid congestion games, and prove that the matroid property is the maximal 
property with respect to the combinatorial structure of the players' strategy 
spaces guaranteeing the existence of Nash equilibria. In such games,
the players' strategy spaces are sets of bases of matroids over the
resources. Gairing, Monien, and Tiemann~\cite{Gairing_etal:ICALP06} consider 
player-specific singleton congestion games with linear delay functions without 
offsets and prove among other results that such games are potential games.

A model closely related to player-specific congestion games are standard 
congestion games. Rosenthal~\cite{Rosenthal:JGame73} proves that these games 
are potential games. Ieong \etal~\cite{Ieong_etal:AAAI05} address the 
convergence time of the best responses dynamics in standard singleton 
congestion games. They show that the best response dynamics converges quickly. 
Fabrikant, Papadimitriou, and Talwar~\cite{Fabrikant_etal:STOC04} show that in 
general standard congestion games players do not convergence quickly. Their 
result holds especially in the case of network congestion games, in which 
players seek to allocate paths in a network. Later, Ackermann, R\"oglin, and 
V\"ocking~\cite{Ackermann_etal:FOCS06} extended the result of Ieong 
\etal~\cite{Ieong_etal:AAAI05} to matroid congestion games, and prove that the 
matroid property is the maximal property on the players' strategy spaces 
guaranteeing polynomial convergence time. Even-Dar 
\etal\cite{Even-Dar_etal:ICALP03} consider the convergence time in standard 
singleton congestion games with weighted players.

Another model which possesses similar properties as player-specific singleton 
congestion games are two-sided markets. In these games, we are given a set of 
resources and a set of players, and for every resource and every player a 
preference list of the elements of the other set. Given such a game, one seeks 
for a stable matching assigning players to resources such that there
exists no pair of player and resource that are not matched to each
other but prefer each other to their current matches.
Gale and 
Shapely~\cite{GaleShapley:AMM62} prove that stable matchings always exist. 
Knuth~\cite{Knuth:Book76} proposes to study better or best response
dynamics in such games and observes that they can cycle. However, Roth
and Vande Vate~\cite{RothVandeVate:Econo90} observe that short better response paths to 
stable matchings always exists. Ackermann \etal~\cite{Ackermann_etal:AIB07} 
follow this line of research and prove an exponential lower bound on the
expected time until the random better (best) response dynamics terminates.
  \section{Player-specific Congestion Games on Trees}

In this section, we consider player-specific congestion games on trees.
Note that in such games the number of resources equals the number of
players. First, we observe that one can always replace the
player-specific delay functions by common delay functions such that the
players' types are preserved. Hence, we obtain a standard singleton
congestion game, whose transition graph equals the transition graph of
the player-specific game. We prove the following theorem.

\begin{theorem} \label{Theorem:Trees}
  In every player-specific congestion game on a tree with $n$ nodes,
  every best response schedule terminates after at most $2n^2$ steps.
\end{theorem}

\begin{proof}
Let $\Gamma$ be a player-specific congestion game $\Gamma$ on a tree. In the 
following, we describe how to replace the player-specific delay functions of 
$\Gamma$ by common delay functions $d_r: \NN \rightarrow \NN$, $r \in \calR$, 
with the following property: For every player $i$ its type with respect to the 
player-specific delay functions equals its type with respect to the standard 
delay functions. Remember that the types completely describe the preferences of 
the players, and hence, the transition graph of $\Gamma$ is not affected by 
replacing the player-specific delay functions by common ones. Since the 
resulting game is a standard singleton congestion game, $\Gamma$ is a 
potential game and we can apply the result of Ieong 
\etal~\cite{Ieong_etal:AAAI05} to upper bound the convergence time. Obviously, 
the same bound holds in $\Gamma$. Thus, by applying the proof of the 
convergence time in standard singleton congestion game as presented 
in~\cite{Ackermann_etal:FOCS06}, we conclude that every best response schedule 
for player-specific congestion games on trees terminates after at most
$2n^2$ steps.

We prove the theorem by induction on the number of players and describe
how to construct a sequence of player-specific congestion games
$\Gamma_1, \ldots, \Gamma_{n}$ on trees with the following properties.
$\Gamma_1$ is obtained from $\Gamma$ by removing the players $2$ to $n$
from the game. The set of resources in $\Gamma_0$ is the set of the two
resources the first player is interested in. Now $\Gamma_i$ is obtained
from $\Gamma_{i-1}$ by adding one player and one resource to $\Gamma_i$.
The player and the resource is chosen in such a way that $\Gamma_i$ is a
player-specific congestion game on a tree. That is, we choose a player
$i$ who is interested in resource $r$ of $\Gamma_{i-1}$, and add
the additional resource $r'$ the player is interested in to $\Gamma_i$.

Obviously $\Gamma_1$, the player-specific congestion game with a single
player and two resources, is a standard congestion game. As induction
hypothesis assume, that we already replaced the player-specific delay
functions in $\Gamma_{i-1}$ by common ones without affecting the players'
types. For ease of notation let $\Gamma_{i-1}^*$ be this game. In the
following, we assume that for every resource $r$ in $\Gamma_i^*$ its
delay functions is defined for all possible congestion values $n_r$
between $1$ and $n$ and not only for the maximum number of players that
are interested in $r$ in $\Gamma_i^*$.

Now given $\Gamma_{i-1}^*$, we describe how to choose the delay functions 
$d_{r}$ of the resources in $\Gamma_i^*$ such that the players in $\Gamma_i^*$ 
and $\Gamma_i$ have the same types. The delay functions of the resources $r$ 
that belong to $\Gamma_{i-1}^*$ are the same as in $\Gamma_{i-1}^*$. 
Additionally, we assume that for every such resource $r$ and every congestion 
$1 < n_r \le n$, $d_r(n_r) - d_r(n_r-1) \ge n$. If this is not the
case, then due to our assumption that the delay functions are strictly increasing, we can 
scale all delays by a factor of $n$ in order to achieve the desired goal. Thus, 
it remains to choose a delay function of the additional resource $r'$ that does 
not belong to $\Gamma_{i-1}^*$. Since the gap between consecutive
values of the delay function $d_r$ is large enough, we can realize
every type for the additional player by choosing the delay function
$d_{r'}$ appropriately. 

Applying the result from~\cite{Ackermann_etal:FOCS06} to the game
$\Gamma_n^*$ directly implies the theorem.
\end{proof}
  \section{Player-specific Congestion Games on Circles}
 
In this section, we consider player-specific congestion games on circles. 
Without loss of generality, we assume that the resources are enumerated from 
$0,\ldots,n-1$, and that they are arranged in increasing order clockwise. 
Furthermore, we assume \wlg that for every player $i$, $\Sigma_i =
\{r_i, r_{i+1\bmod n}\}$. In the following, we call $r_i$ the 0- and $r_{i+1\bmod n}$ 
the 1-strategy of player $i$. Furthermore, we drop the \hspace{-1.5ex}$\mod n$ 
terms and assume that all indices are computed modulo $n$. Due to our 
assumptions on the delay functions, there are six different types of players in 
such games:

\[\begin{array}{lclclclr} d_{r_i}^i(1) & < & d_{r_i}^i(2) & < & 
d_{r_{i+1}}^i(1) & < & d_{r_{i+1}}^i(2) & \mbox{\hspace{5ex} type 1} \\[1ex] 
d_{r_i}^i(1) & < & d_{r_{i+1}}^i(1) & < & d_{r_i}^i(2) & < & d_{r_{i+1}}^i(2) & 
\mbox{\hspace{5ex} type 2} \\[1ex] d_{r_i}^i(1) & < &   d_{r_{i+1}}^i(1) & < & 
d_{r_{i+1}}^i(2) & < & d_{r_i}^i(2) & \mbox{\hspace{5ex} type 3}  \end{array}\]

We call the three other types, which can be obtained by exchanging the
identities of the resources $r_i$ and $r_{i+1}$ in the above
inequalities, type $1'$, type $2'$, and type $3'$. Furthermore, we call
two players $i$ and $j$ \emph{consecutive}, if they share a resource,
that is, if $j=i+1$ or $i=j+1$. Given a state $S$, we call two
consecutive players \emph{synchronized}, if both play the same strategy,
that is, if both either play their 0- or their 1-strategy. Moreover, we
call a set of consecutive players $i, \ldots, j$ synchronized if all
players play the same strategy.

\subsection{Cycles in the Transition Graphs and a Lower Bound}

We present an infinite family of games possessing cycles in their transition 
graphs. From this construction we derive a lower bound of $\Omega(n^2)$ on the 
convergence time of the random best response dynamics in player-specific 
congestion games on circles.

Consider a game on a circle with $n$ players which are all of type 3.
It is not difficult to verify that this game possesses only two Nash equilibria: either 
all players play their 0-strategy or their 1-strategy. Consider now a state $S$ 
with the following properties: In $S$ we can partition the players into two 
non-empty sets ${\cal S}_0$ and ${\cal S}_1$ of synchronized players. Players 
in ${\cal S}_0$ all play their 0-strategy, whereas players in ${\cal S}_1$ all 
play their 1-strategy. Again, it is not difficult to verify that in every such 
state there are exactly two players who have an incentive to change their 
strategies. From both sets only the first player clockwise has an incentive to 
change its strategy. Thus, there exist cycles in the transition graphs of these 
games. We obtain such a cycle by selecting players from the two sets 
alternately, and letting them play best responses.

In order to prove a lower bound on the convergence time of the random best 
response dynamics, observe that with probability $1/2$ the total number of 
players playing their 0-strategy increases or decreases by one whenever a 
player is selected uniformly at random. After the strategy change either all 
players are synchronized, and therefore the random best response dynamics 
terminates, or again we are in a state $S'$ with two sets of synchronized 
players. Observe now that this process is isomorphic to a random walk on a line 
with nodes $v_0, \ldots, v_n$. The node $v_i$ corresponds to the fact that $i$ 
players play their 0-strategy. As the expected time of a random walk on a line 
with $n+1$ nodes to reach one of the two ends of the line is $\Theta(n^2)$ if 
the walk starts in the middle of the line~\cite{Lovasz:Book96}, we obtain a 
lower bound of $\Omega(n^2)$.

\begin{corollary}\label{Corollaray:PS_Circle_LowerBound}
  There exists an infinite family of instances of player-specific congestion 
  games on circles with corresponding initial states such that the number of 
  steps until the random best response dynamics terminates is lower bounded by 
  $\Omega(n^2)$.
\end{corollary}

\subsection{An Upper Bound}

In this section, we present an upper bound on the convergence time of the 
random best response dynamics in player-specific congestion games on circles. 
We prove the following theorem which matches the lower bound presented in 
Corollary~\ref{Corollaray:PS_Circle_LowerBound}.

\begin{theorem}
  In every player-specific congestion game on a circle the random best response 
  dynamics terminates after $O(n^2)$ steps in expectation.
\end{theorem}

The remainder of this section is organized as follows. We characterize with 
respect to the types of the players in which cases there are cycles in the 
transition graphs of such games. We show that cycles only exist if all players 
are of type 3 or type $3'$. We analyze the convergence time of deterministic 
best response dynamics in games with acyclic transition graphs by developing a 
general framework that allows to derive potential functions from which one can 
easily derive upper bounds. Finally, we analyze the convergence time of the 
random best response dynamics in the case of games with players of type 3 or 
type $3'$.

\subsubsection{The Impact of Type 1 Players}

First, we investigate the impact of type 1 players on the existence of cycles 
in the transition graphs and on the convergence time of the best response 
dynamics. We claim that games with at least one player of type 1 do not possess 
cycles in their transition graphs. Intuitively, this is true since every player 
of type 1 changes its strategy at most once, whereas in a cycle every player 
changes its strategy at least twice.

\begin{lemma}\label{Lemma:Type1}
  Let $\Gamma$ be a player-specific congestion game on a circle. If there 
  exists at least one player of type 1 or $1'$, then $TG(\Gamma)$ is acyclic. 
  Moreover, every best response schedule terminates after at most $4n^2$ steps.
\end{lemma}

In order to prove Lemma~\ref{Lemma:Type1}, we first prove the following one.

\begin{lemma} \label{Lemma:Cycles}
  Let $\Gamma$ be a player-specific congestion game on a circle whose 
  transition graph contains cycles. Then \emph{every} player changes its 
  strategy at least twice in every cycle of $TG(\Gamma)$.
\end{lemma}

\begin{proof}
The fact that players being involved in the cycle change their strategy an even 
number of times is obvious. Thus, it remains to show that \emph{every} player 
changes its strategy. For contradiction, assume that there exists a player $i$ 
and a cycle in $TG(\Gamma)$ such that player $i$ does not change its strategy 
on that cycle. In this case, we could remove the player from the game, and 
artificially increase the congestion on the resource the player allocates by 
one. We would then obtain a player-specific congestion game on a tree which 
cannot have a cycle in the transition graph due to Theorem~\ref{Theorem:Trees}.
\end{proof}

Next we prove Lemma~\ref{Lemma:Type1} for type 1 players.
The proof for type $1'$ players is essentially the same.

\begin{proof}[Proof of Lemma~\ref{Lemma:Type1}] Without loss of generality, let 
player $0$ be of type 1. Then observe that player $0$ will never play its 
1-strategy again, once it played its 0-strategy. Thus, by 
Lemma~\ref{Lemma:Cycles}, $TG(\Gamma)$ is acyclic.

In order to prove the convergence time, observe that if we fix player 0 to one 
of its strategies, then we obtain a player-specific congestion game on a tree. 
Due to Theorem~\ref{Theorem:Trees}, the convergence time of such games is upper 
bounded by $2n^2$. Furthermore, observe that the transition graphs of 
these games are isomorphic to disjoint subgraphs of $TG(\Gamma)$. The first 
subgraph contains all nodes of $TG(\Gamma)$ in which player 0 plays its 
0-strategy, the second one contains all nodes in which player 0 plays
its 1-strategy. Finally, as all edges between these two subgraph are directed from 
the second one to the first one, and as the maximal length of any best response 
sequence in each of these subgraphs is upper bounded by $2n^2$, the
lemma follows.
\end{proof}

In the following, we will assume that there exists no player of type 1 or $1'$, 
as otherwise we could apply Lemma~\ref{Lemma:Type1}.

\subsection{A Framework to Analyze the Convergence Time}

In this section, we present a framework to analyze the convergence time of best 
response schedules in player-specific congestion games on circles. Let $\Gamma$ 
be a game such that there is no player of type 1 or $1'$. First, we investigate 
whether there is a sufficient condition such that player $i$ does not want to 
change its strategy in a state $S$ of $\Gamma$.

\begin{observation} \label{Observation:Synch1}
  Suppose that player $i$ is not of type $1$ or $1'$. Then if it is 
  synchronized with the players $i-1$ and $i+1$ in $S$, it has no incentive to 
  change its strategy.
\end{observation}

In the following, we call a resource $r$ \emph{overloaded} in state $S$
if two players share $r$. Additionally, we call a resource $r'$
\emph{underloaded} in state $S$ if no player allocates $r'$. Obviously
in every state of $\Gamma$, the total number of overloaded resources
equals the total number of underloaded resources. From
Observation~\ref{Observation:Synch1}, we conclude that in every state $S$
only players who allocate a resource that is currently overloaded or who
could allocate a resource that is currently underloaded might have an
incentive to change their strategy.

Based on this observation, we now present a general framework to analyze the 
convergence time of best response schedules. First, we introduce the notion of 
\emph{over- and underload tokens}. Given an arbitrary state $S$ of $\Gamma$, 
we place an \emph{overload token} on every overloaded resource. Additionally, 
we place an \emph{underloaded token} on every underloaded resource. Obviously 
over- and underload tokens alternate on the circle. Furthermore, note that a 
legal placement of tokens uniquely determines the strategies the players play. 
A placement of tokens is legal if no two tokens share a resource, and if the 
tokens alternate on the circle.

\begin{figure}[ht]\begin{center}

  \psfrag{overloaded}{overloaded}
  \psfrag{underloaded}{underloaded}
  \psfrag{orientation of the players}{orientation of the players}
  \psfrag{ri}{$r_i$}
  \psfrag{rj}{$r_j$} 
  
  \psfrag{table}{ \begin{tabular}{|l|l|l|} \hline & overload      & underload 
  \\ \hline type 2  & anticlockwise &     clockwise \\
  type $2'$ &     clockwise & anticlockwise \\
  type 3  &     clockwise &     clockwise \\
  type $3'$ & anticlockwise & anticlockwise \\ \hline
  \end{tabular}}

  \includegraphics{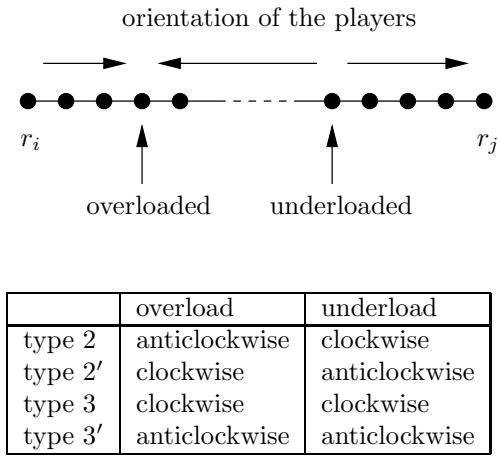}                  
  \caption{In which directions do the tokens move?}
  \label{Fig:Move}
\end{center}\end{figure}

In the following, we investigate in which directions tokens move if players 
play best responses. Consider first a sequence of resources $r_i, \ldots, r_j$ 
and assume that players $i, \ldots, j-1$ are of the same type $t$. 
Additionally, assume that an overload token is placed on resource $r_k$, and 
that an underload token is placed on resource $r_l$ with $i < k < l < j$. The 
scenario we consider is depicted in Figure~\ref{Fig:Move}.

Assume first, that the distance (number of edges) between the two tokens
is at least two, i.e., $|l-k|\ge2$. In this case, each token can only
move in one direction. The directions are uniquely determined by the type
of the players. They can be derived from investigating, with respect to
the players' type $t$, which players have incentives to change their
strategy. The directions are stated in Figure~\ref{Fig:Move}, too. Assume
now that the distance between the two tokens is one. That is, $k=l-1$.
Then, there exists a player who is interested in the over- and
underloaded resource, and who currently allocates the overloaded one. It
is not difficult to verify that this player always has an incentive to
change its strategy. Note that this holds regardless of the player's type
since we assumed that there are no players of type 1 and $1'$. Observe
that after the strategy change of this player all players $i,\ldots,j-1$
are synchronized and therefore there exist no over- and underloaded
resources anymore. In the following, we call such an event a
\emph{collision of tokens}.

So far, we considered sequences of players of the same type and observed
that there is a unique direction in which tokens of the same kind move.
In sequences with multiple types of players such unique directions do not
exist any longer, i.e., overload as well as underload tokens can move in
both directions. However, if two players of different types share a
resource and if due to best responses of both players an over- or
underload token moves onto this resource, then the token could stop
there. In the following, we formalize this observation with respect to
overload tokens and introduce the notion of \emph{termination points}.

\begin{definition}
We call a resource $r_i$ a \emph{termination point} of an overload token if 
the following conditions are satisfied.
\begin{enumerate}
  \item The players $i-1$ and $i$ have different types. Let these types be 
  $t_{i-1}$ and $t_i$.
  \item In sets of consecutive players of type $t_{i-1}$ overload tokens move 
  clockwise, whereas they move anticlockwise in sets of consecutive players of 
  type $t_i$.
\end{enumerate}
\end{definition}

\begin{figure}[ht] \centering
\subfigure[Example of a termination point\label{Fig:TPointsA}]{
  \psfrag{type2}{type 2}
  \psfrag{type3}{type 3}
  \psfrag{overloaded}{overloaded}
  \psfrag{ri}{$r_i$}
  \psfrag{orientation of the players}{orientation of the players}
  \includegraphics{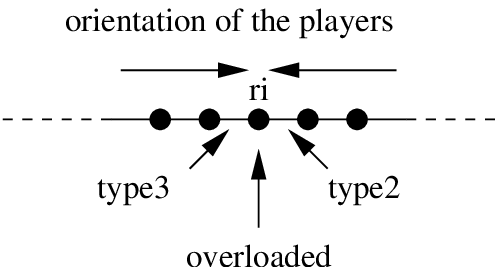}
}
\hspace{0.5cm}
\subtable[List of all termination points\label{Fig:TPointsB}]{
  \hspace{0.1cm}

  \begin{tabular}[b]{|c|c|} \hline $ \rightarrow \rightarrow 
  \leftarrow \leftarrow $ & $\leftarrow \leftarrow \rightarrow
  \rightarrow$ \\\hline $2'$\:  2 & 2\:\: $2'$ \\
  3\:\: $3'$ & 3\:\: $3'$ \\
  3\:\: 2 & -        \\
  -       & 2\:\: $3'$ \\
  -       & 3\:\: $2'$ \\
  $2'$\: $3'$ & - \\ \hline
  \end{tabular}\hspace{0.1cm}
  \vspace{0.1cm}
}
\end{figure}

We illustrate the definition in Figure~\ref{Fig:TPointsA}. Let player
$i-1$ be of type 3, and let player $i$ be of type 2. In this case, the requirements of 
the definition are satisfied. Assume, that player $i-1$ plays its 1-strategy 
and that it is synchronized with player $i-2$. Additionally, assume that player 
$i$ plays its 0-strategy and that it is synchronized with player $i+1$. Observe 
now that the token cannot move as neither player $i-1$ nor player $i$ has an 
incentive to change its strategy. Suppose now that initially all players along 
the path play their 0-strategy. Then an overload token that moves from the left 
to the right along the path stops at $r_i$. The token may only move on if one 
of the two players is not synchronized with its neighbor any longer. In this 
case, this player always has an incentive to change its strategy as it can 
allocate a resource that is currently underloaded. Thus, an underload and an 
overload token collide. Additionally, if initially all players play their 
1-strategy and an overload token moves from the right to the left along the 
path, we observe the same phenomenon. The token cannot pass the resource $r_i$ 
unless it collides with an underload token.

Note that the definition of a termination point can easily be adopted to 
underload tokens. A list of all termination points is given in 
Figure~\ref{Fig:TPointsB}. In the left column we present all
termination points for overload tokens, in the right one for underload tokens.

\subsection{Analyzing the Convergence Time}

In this section, we analyze the convergence time in player-specific congestion 
games on circles. We distinguish between the following four cases.

\begin{description}
  \item[\hspace*{2ex} Case 1:] For both kinds of tokens there exists at least 
  one termination point.
  \item[\hspace*{2ex} Case 2:] Only for one kind of tokens there exists at 
  least one termination point.
  \item[\hspace*{2ex} Case 3:] There exist no termination points but over- and 
  underload tokens move in opposite directions.
  \item[\hspace*{2ex} Case 4:] There exist no termination points and over- and 
  underload tokens move in the same direction.
\end{description} 

In the first two cases, we present potential functions and prove that the 
transition graphs of such games are acyclic and that every best response 
schedule terminates after $O(n^2)$ steps. In the third case, we can do slightly 
better and prove an upper bound of $O(n)$ on the convergence time. In all cases 
one can easily construct matching lower bounds. Only in the fourth case 
deterministic best response schedules can cycle. In this case, we prove that 
the random best response schedule terminates after $O(n^2)$ steps in 
expectation.
  
Before we take a closer look at the different cases, we discuss which games 
with respect to their players' types belong to which case. Games only with 
players of type 2 and $2'$ or only with players of type 3 and $3'$ belong to 
the first case. Additionally, some games with more than two types of players 
belong to this case. The second case covers all games with at least three 
different kinds of players which do not belong to the first case. Furthermore, 
it covers games with type 2 and type 3 players, with type $2'$ and type $3'$ 
players, type $2'$ and type 3 players, and with type 2 and type $3'$ players. 
Games with type 2 players only, or games with type $2'$ players only belong to 
the third case. Finally, games with type 3 players only and games with type 
$3'$ players only belong to the fourth case. These observations can easily be 
derived from Figure~\ref{Fig:TPointsB}.

\subsubsection{Case 1}

\begin{lemma} \label{Lemma:2TerminationPoints}
  Let $\Gamma$ be a player-specific congestion game on a circle with
  termination points for both kinds of tokens. Then $\Gamma$ is a
  potential game, and every best response schedule terminates after $O(n^2)$ steps.
\end{lemma}

\begin{proof} Let $S$ be a state of $\Gamma$ and consider the mapping that maps 
every token in $S$ to the next termination point lying in the direction in 
which the token moves. In the following, we define $d(t,S)$ as the distance of 
a token $t$ in state $S$ to its corresponding termination point. Obviously 
$d(t,S) \le n$. Consider now the potential function $\phi(S) = 
\sum_{\text{\small token } t} d(t,S)$ and suppose that a player plays a best 
response. Then either one token moves closer to its termination point or two 
tokens collide. In both cases $\phi(S)$ decreases by at least 1. Thus, 
$\phi(S)$ strictly decreases if a player plays a best response and therefore, 
$TG(\Gamma)$ is acyclic. Moreover, as $\phi(S)$ is upper bounded by $O(n^2)$, 
every best response schedule terminates after $O(n^2)$ steps.
\end{proof}

\subsubsection{Case 2}

\begin{lemma} \label{Lemma:1TerminationPoint}
  Let $\Gamma$ be a player-specific congestion game on a circle with 
  termination points only for one kind of token. Then $\Gamma$ is a 
  potential game, and every best response schedule terminates after $O(n^2)$ 
  steps.
\end{lemma}

\begin{proof}
Without loss of generality, assume that termination points only exist for 
overload tokens. In this case, we define $d(t_o,S)$ for every overload token 
$t_o$ as in the proof of Lemma~\ref{Lemma:2TerminationPoints}. For every 
underload token $t_u$ we define $d(t_u,S)$ as follows. Let $t_o$ be the first 
overload token lying in the same direction as $t_u$ moves.

\begin{enumerate}
  \item If $t_o$ moves in the opposite direction than $t_u$, then we define 
  $d(t_u,S)$ as the distance between the two tokens. The distance of two tokens 
  moving in opposite directions is defined as the number of moves of these 
  tokens until they collide.
  \item If $t_o$ moves in the same directions as $t_u$ then we define 
  $d(t_u,S)$ as the distance between $t_u$ and $t_o$ plus the distance between 
  $t_o$ and the first termination point at which $t_o$ has to stop. Thus, 
  $d(t_u,S)$ equals the maximum number of moves of these two tokens until they 
  collide.
\end{enumerate}

Observe, that for every underload token $t_u$, $d(t_u,S) \le 2n$.
Consider, the potential function $\phi \colon \Sigma_1 \times \ldots \times \Sigma_n 
\rightarrow \NN \times \NN$ with $\phi(S) = (\phi_1(S), \phi_2(S))$, where 
$\phi_1(S)$ equals the total number of overload tokens in $S$ and $\phi_2(S)$ 
equals the sum of all $d(t,S)$ for all under- and overload tokens. Suppose now 
that a player plays a best response. Obviously if two tokens collide, then 
$\phi_1(S)$ decrease by one. Moreover, if there is no collision, then 
$\phi_2(S)$ decreases. Note that in the first case $\phi_2$ may increase. This 
may happen if, due to the collision, $d(t_u,S)$ of a remaining underload token 
$t_u$ has to be recomputed as its associated overload token has been removed. 
The new value is upper bounded by the sum of the old values of $t_u$ and the 
collided underload token plus 1. Now consider the lexicographic ordering 
$<_{\phi}$ of the states of $\Gamma$ with respect to $\phi$. Let $S$ and $S'$ 
be two states of $\Gamma$. Then
\begin{eqnarray*} 
S <_{\phi} S' & \Leftrightarrow & \left \{\begin{array}{llll} \phi_1(S) < 
\phi_1(S') & & & \mbox{or } \\
\phi_1(S) = \phi_1(S') & \mbox{ and } & \phi_2(S) < \phi_2(S') \enspace . 
\end{array} \right .
\end{eqnarray*}
Observe that $\phi$ strictly decreases if a player plays a best response. Thus, 
$TG(\Gamma)$ is acyclic. Additionally, observe that $\phi_1$ is upper bounded 
by $n$, and that $\phi_2$ is upper bounded by $n^2$. However, as $\phi_2$ only 
increases by one when $\phi_1$ decreases, we conclude that every best response 
schedule terminates after $O(n^2)$ steps.
\end{proof}

\subsubsection{Case 3}  

\begin{lemma} \label{Lemma:NoTerminationPoint}
  Let $\Gamma$ be a player-specific congestion game on a circle with
  no termination points in which over- and underload tokens move in
  opposite directions. Then $\Gamma$ is a potential game, and every best response 
  schedule terminates after $O(n)$ steps.
\end{lemma}

\begin{proof}
Let $S$ be a state of $\Gamma$ and consider the one-to-one mapping that maps 
every overload token to the next underload token lying in the direction in 
which the token moves. We define the distance of such a pair of tokens as the 
maximum number of moves of these two tokens till they collide.

Suppose now that a player plays a best response. Then either the number of 
overload tokens or the distance between one pair of tokens decreases by one. 
Consider now the potential function $\phi \colon \Sigma \rightarrow \NN \times 
\NN$ with $\phi(S) = (\phi_1(S), \phi_2(S))$, where $\phi_1(S)$ equals the 
number of overload tokens in $S$, and $\phi_2(S)$ equals the sum of all 
distances of pairs of tokens. Observe now that in the case of a best response, 
$\phi_1$ either decreases by 1 or remains unchanged. In the first case, 
$\phi_2$ may increase by 1. This is true as tokens from different pairs may 
collide. However, this can happen at most $n$ times. If this happens, the 
remaining two tokens form a new pair whose distance equals the sum of the 
distances of the previous pairs plus 1. In the second case, $\phi_2$ decreases 
by 1. Then by similar arguments as in the proof of 
Theorem~\ref{Lemma:1TerminationPoint}, we conclude that $TG(\Gamma)$ is 
acyclic. Finally, observe that $\phi_1$ is upper bounded by $n$. Moreover, 
$\phi_2$ is upper bounded by $n$, too. Finally, as $\phi_2$ only increases by 
one when $\phi_1$ decreases, we conclude that every best response schedule 
terminates after $O(n)$ steps.
\end{proof}
\subsubsection{Case 4}
In the following, we present a proof of the fourth case for players 
of type 3. By symmetry of the types 3 and $3'$, the same result holds
for games with players of type $3'$, too.

\begin{lemma} \label{Lemma:Type3}
  Let $\Gamma$ be a player-specific congestion game on a circle in
  which all players are of type 3. Then the random best response
  schedule terminates after $O(n^2)$ steps in expectation.
\end{lemma}

\begin{proof} In order to prove the lemma, we prove the following lemma.

\begin{lemma}\label{Lemma:Type3Equal}
In every state $S$ of $\Gamma$ the number of players who want to change from 
their 0- to their 1-strategy equals the number of players who want to change 
from their 1- to their 0-strategy.
\end{lemma}
 
\begin{proof}
In the following, we call a synchronized set of consecutive players
\emph{maximal} if the next players to both ends of the set play different
strategies than the synchronized players. Obviously in every state $S$ of
$\Gamma$ which is not an equilibrium the number of maximal synchronized
sets of players playing their 0-strategy equals the number of maximal
synchronized sets of players playing their 1-strategy.

We now prove that in every maximal synchronized set of consecutive players only 
the first player clockwise has an incentive to change its strategy. Thus, in 
every maximal set, there is only a single player who wants to change its 
strategy. Note that this suffices to prove the lemma.

First, consider a maximal, synchronized subset of consecutive players $\calN' = 
\{i, \ldots $ $j\}$ which all play their 0-strategy. Then player $i-1$ 
plays its 1-strategy, and therefore the players $i-1$ and $i$ share 
resource $r_i$. In this case, player $i$ can decrease its delay by changing to 
her 1-strategy. Other players $k \in \calN'$, $k\neq i$, do not have an 
incentive to change their strategy as this would increase their delay.

Second, consider a maximal synchronized subset of consecutive players $\calN' = 
\{i, \ldots j\}$ which all play their 1-strategy. Then player $i-1$ plays its 
0-strategy and therefore no player currently allocates resource $r_i$. 
Observe now that player $i$ may decrease its delay by changing to its 
0-strategy. Again, all other players $k \in \calN'$, $k\neq i$, do not have an 
incentive to change their strategy as this would increase their delay. This is 
especially true for the last player, who currently allocates an overloaded 
resource.
\end{proof}  

Consider now the random best response schedule activating an unsatisfied player 
uniformly at random. From Lemma~\ref{Lemma:Type3Equal} we conclude that the 
total number of players playing their 0-strategy increases or decreases by 1 
with probability $1/2$. Combining this with the observation that in a
Nash equilibrium all players play the same strategy, we conclude that the 
random best response schedule is isomorphic to a random walk on a line with 
$n+1$ vertices. Vertex $v_i$ corresponds to the fact that $i$ players play 
their 0-strategy. As the time of such a random walk to reach one of the two 
ends of the line is $O(n^2)$, the lemma follows.
\end{proof}
\section{Player-specific Congestion Games on General Graphs}

In this section, we consider player-specific congestion games on general
graphs and present evidence supporting
Conjecture~\ref{Conjecture:PS_lowerBound} by constructing a family of
instances for which experimental results clearly show a super-polynomial
convergence time. Our analysis of player-specific congestion games on
circles is based on the notion of over- and underload tokens, and there
is no straightforward extension of this notion to player-specific
singleton congestion games on general graphs. The instances we construct
have, however, the property that every resource has a fixed congestion
that is taken in every Nash equilibrium, and we can define tokens with
respect to these congestions. To be precise, if the congestion on a
resource deviates by $x$ from the equilibrium congestion, we place $x$
overload tokens in the case $x>0$ and we place $-x$ underload tokens in
the case $x<0$. Note that for circles with type 3 players this definition
coincides with the former definition of tokens.

The crucial property of games on circles with type 3 players leading to
polynomial convergence is that the number of tokens cannot increase. The
instances we construct in this section are in some sense similar to
circles with type 3 players, but we attach additional gadgets to the
nodes which can occasionally increase the number of tokens. We start with
a circle with $n$ type 3 players and replace each edge by $n$ parallel
edges. This modification allows each node to store more than one token of
the same kind if the preferences of the players are adjusted accordingly.
Other properties are not affected by this modification, that is, over-
and underload tokens still move in the same direction with approximately
the same speed and if an overload and an underload token meet, they both
vanish. Each time a node contains at least two tokens of the same kind,
the gadget attached to the node is triggered with constant probability.
If a gadget is triggered, it can emit a new pair of overload and
underload token into the circle. Usually, this new pair is stored in
the gadget and only emitted after the triggering tokens have moved on
a linear number of steps. The new tokens are not emitted simultaneously
but the second is usually only emitted after the first one has moved on
a linear number of steps in order to prevent the new tokens from
canceling each other out immediately.

Initially we introduce two overload tokens at node $0$ and two underload
tokens at node $n/2$. The two overload tokens move independently through
the circle starting at the same node. Typically they meet a couple of
times before they meet the underload tokens and vanish. The same is true
for the underload tokens as well, meaning that typically a couple of
gadgets get triggered before the initial tokens vanish. Hence, the number
of tokens has a tendency to increase. Since the triggered gadgets emit
the stored tokens in a random order, the random process soon becomes
unwieldy and we fail to rigorously prove that it takes super-polynomial
time in expectation until all tokens vanish. This conjecture is, however,
strongly supported by simulations.

\subsection{Our Construction}
\begin{figure}[ht] \begin{center}
\subfigure[Gadget $G_i$ \label{Fig:Gadget_lowerBound}]{
  \psfrag{e0}{$e_{i,0}$}
  \psfrag{e1}{$e_{i,1}$}
  \psfrag{e2}{$e_{i,2}$}
  \psfrag{e3}{$e_{i,3}$}
  \psfrag{e4}{$e_{i,4}$}
  \psfrag{r0}{$r_{i,0}$}
  \psfrag{r1}{$r_{i,1}$}
  \psfrag{r2}{$r_{i,3}$}
  \psfrag{r3}{$r_{i,2}$}  
\includegraphics[scale=1.0]{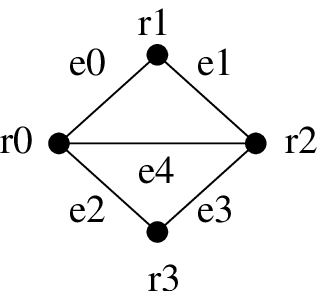} } \qquad \subfigure[The 
player-specific delay functions \label{Fig:Thresholds}]{
  \psfrag{table}{\begin{tabular}{lll}
type 0 & & $t_0 = 3n$ \\
type 1 & & $t_1 = n - 1$ \\
type 2 & & $t_2 = 3n - 2$ \\
type 3 & & $t_3 = n - 1$ \\
type 4 & & $t_4 = 3n - 1$ \\
  \end{tabular}}
\includegraphics[scale=1.0]{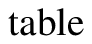} }
\caption{The lower bound construction}
\end{center}\end{figure}

Given $n \in \NN$ we construct a player-specific congestion game $\Gamma_n$ 
consisting of $n$ gadgets $G_0, \ldots, G_{n-1}$ as follows. In the following, 
the notion of a gadget differs from the notion used in the previous discussion. 
Previously, we described how to attach gadgets to a circle in order to 
illustrate the relation to games on circles. Next gadgets are arranged on a 
circle. A single gadget $G_i$ is depicted in 
Figure~\ref{Fig:Gadget_lowerBound}. It consists of $4$ resources 
$r_{i,0},\ldots,r_{i,3}$ and $5n$ players. Each edge in the figure represents 
$n$ of them. The gadgets are arranged on a circle, such that for every $i$ the 
resources $r_{i,3}$ and $r_{i+1,0}$ coincide. Thus, for every $i$, $6n$ players 
are interested in $r_{i,0}$ and $r_{i,3}$, and $2n$ players are interested in 
$r_{i,1}$ and $r_{i,2}$.

For every player who chooses between the two resources $r_{i,k}$ and $r_{i,l}$ 
with $l<k$ we call $r_{i,l}$ the 0-strategy and $r_{i,k}$ the 1-strategy of 
that player. In the following, we refer by the term \emph{type $j$ player} to a 
player represented by an edge $e_{i,j}$. The player-specific delay functions 
are defined as follows. All players of the same type $j$ have the same 
functions for the two resources they choose between. We define these functions 
in terms of a threshold $t$ for their 0-strategies, meaning that the 0-strategy 
is a best response if and only if the total number of \emph{other} players 
allocating the 0-strategy resource is less or equal to the threshold $t$. 
Otherwise the 1-strategy is the best response. The thresholds are defined as 
depicted in Figure~\ref{Fig:Thresholds}.

In the next sections, we prove that every resource has the same congestion in 
every Nash equilibrium. We proceed with a description of how gadgets can 
generate new tokens. Finally, we present results obtained from simulations.

\subsection{Properties of Nash Equilibria}

In order to simplify our proceeding discussion, we
introduce the term $c_{i,j}^b(S) \in \NN$, $b \in \{0,1\}$, to denote the number of type 
$j$ players in gadget $i$ who play their $b$-strategy in state $S$. 
Furthermore, we define $n_{i,j}(S) = n_{r_{i,j}}(S)$. In the following, let 
$S^*$ be a Nash equilibrium of $\Gamma_n$. For ease of notation, we use 
$c_{i,j}^b = c_{i,j}^b(S^*)$ and $n_{i,j} = n_{i,j}(S^*)$. The following 
observation is true because $S^*$ is a Nash equilibrium.

\begin{observation} \label{Observation:Determined}
  Let $j \in \{1,3\}$ and $b \in \{0,1\}$. Then for every $0 \le i < n$ the 
  number of type $j$-players playing their $b$-strategy in gadget $G_i$ in 
  $S^*$ is uniquely determined by the number of type $j-1$ players playing 
  their $b$-strategy in gadget $G_i$ in $S^*$, i.e., $c_{i,j-1}^b = c_{i,j}^b$.
\end{observation}
 Next, we prove that every resource has the same congestion 
in every Nash equilibrium.

\begin{lemma}  \label{Lemma:UniqueCongestion}
  For every Nash equilibrium $S^*$ of\/ $\Gamma_n$ and every $0 \le i <n$, 
  \[n_{i,0} = 3 \cdot n \mbox{ \,\,and\,\, } n_{i,1} = n_{i,2} = n \enspace .\]
\end{lemma}

\begin{proof}
  First observe that for every gadget $G_i$, it holds \[c_{i,0}^0 \ge
  c_{i,4}^0 \ge c_{i,2}^0 \enspace .\] If the first inequality were not true, then there 
  exist type 0 players in $G_i$ playing their 1-strategy and type 4 players 
  playing their 0-strategy. However, since $S^*$ is a Nash equilibrium,
  all type 4 players in $G_i$ who play their 0-strategy are satisfied and thus $n_{i,0} 
  \le 3n$. We observe that all type 0 players currently playing their 
  1-strategy have an incentive to change their strategy. A similar 
  argument proves the second inequality. Essentially, the same arguments prove 
  the following implications:
  \begin{eqnarray*}
  c_{i,0}^0 < n & \Rightarrow & c_{i,4}^0 = 0, \\
  c_{i,4}^0 < n & \Rightarrow & c_{i,2}^0 = 0.
  \end{eqnarray*}
  Now consider an arbitrary gadget $G_i$ and let $3n - k_{i-1}$ be the number 
  of players from gadget $G_{i-1}$ allocating resource $r_{i,0}$. In the 
  following, we discuss how the parameter $k_{i-1}$ affects the choices of the 
  players in gadget $G_i$ in the Nash equilibrium $S^*$. We prove that the best 
  responses of the players in $G_i$ are uniquely determined by the parameter 
  $k_{i-1}$. In order to do so, we distinguish 6 cases.
  
\begin{description}
  \item[1. Case ${k_{i-1} = 0}$:] All type 1, type 3, and type 4 players in 
  gadget $G_{i-1}$ play their 1-strategy. Due to 
  Observation~\ref{Observation:Determined} we conclude that all type 0 and type 
  2 players in $G_{i-1}$ play their 1-strategy as well, and therefore the 
  congestion on $r_{i-1,0}$ is at most $3n$. In this case, however, all type 0 
  players in $G_{i-1}$ have an incentive to play a best response. We conclude 
  that this case does not appear in a Nash equilibrium.
  \item[2. Case ${1 \le k_{i-1} < n}$:] $k_{i-1}+1$ type 0 and $k_{i-1}+1$ type 
  1 players in $G_i$ play their $0$-strategy. The remaining players in $G_i$ 
  play their $1$-strategy. Thus $k_i = k_{i-1}+1$.
  \item[3. Case ${k_{i-1}=n}$:] All type 0 and all type 1 players in $G_i$ play 
  their 0-strategy; all other players in $G_i$ play their $1$-strategy. Thus
  $k_i = k_{i-1}$.
  \item[4. Case ${n < k_{i-1} \le 2n}$:] All type 0 and all type 1
  players in $G_i$ play their 0-strategy. Additionally, $k_{i-1}-n$ type 4 players in 
  $G_i$ play their $0$-strategy. The remaining players in $G_i$ play their 
  $1$-strategy. Thus $k_{i} = k_{i-1}$.
  \item[5. Case ${2n < k_{i-1} < 3n}$:] All type 0, all type 1 and all type 4 
  players in $G_i$ play their 1-strategy. Additionally, $k_{i-1}-2n-1$ type 3 
  and $k_{i-1}-2n-1$ type 4 players in $G_i$ play their 0-strategy. The 
  remaining players in $G_i$ play their 1-strategy. Thus $k_i = k_{i-1}-1$.
  \item[6. Case ${k_{i-1} = 3n}$:] Similar arguments as in the first case show 
  that this case does not appear in a Nash equilibrium.
\end{description}
  
As an intermediate observation we conclude that the lemma is true if at least 
one gadget $G_i$ exists for which $n \le k_i \le 2n$ holds. In this case 
$k_{i-1} = k_i$ for every $1 \le i <n$ and the players play the strategies as 
described above.

Next we take a closer look at the second and fifth case. We begin with the 
second one in which $1 \le k_{i-1} < n$ implies $k_i = k_{i-1}+1$ which implies 
$k_{i+1} = k_{i-1}+2$ and so on until $k_j=n$. In this case we enter the third 
case which implies $k_{j+1} = n$ and so on. Obviously this leads to a 
contradiction since $k_{i-1} < n$. Thus, whenever there exists a gadget for 
which $k_{i-1} < n$ holds, $S^*$ is not a Nash equilibrium. Similar arguments 
show that the fifth case does not appear in a Nash equilibrium either.
\end{proof}

\subsection{Generating New Tokens}

Motivated by Lemma~\ref{Lemma:UniqueCongestion} we are now ready to introduce a 
new notion of tokens.

\begin{definition}
  Let $S$ be an arbitrary state of $\Gamma_n$ and let $n_r^*$ be the 
  congestion on a resource $r$ in every Nash equilibrium. Then, we place
  over- and underload on the resources according to the following rules.
  \begin{enumerate}
    \item If $n_r(S) = n_r^*+k$, $k \in  \NN$, then we place $k$ overload 
    tokens on $r$.
    \item If $n_r(S) = n_r^*-k$, $k \in  \NN$, then we place $k$ underload 
    tokens on $r$.
  \end{enumerate}
\end{definition}

\begin{figure}[ht]
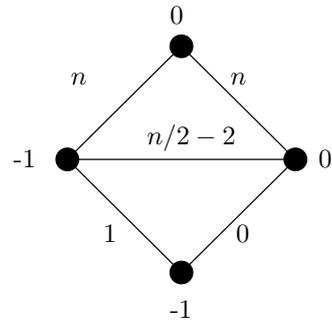
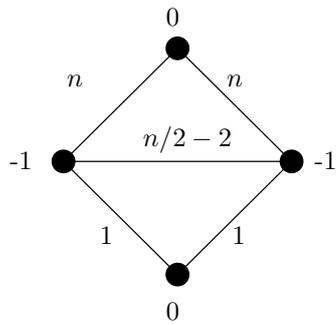
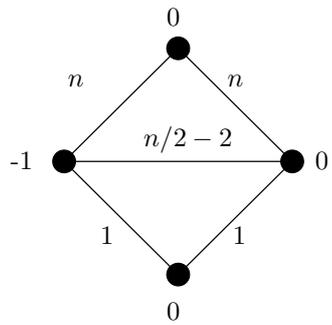
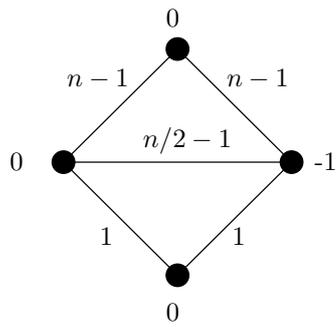
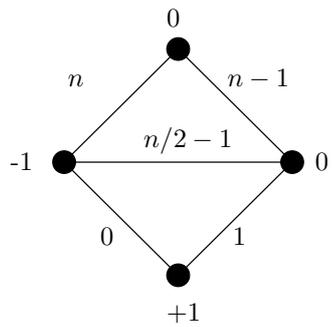
 \begin{center}
\subfigure[Initial configuration.\label{SubFig:A2}]{
  \psfrag{(a,b)_0}{$n$}
  \psfrag{(a,b)_1}{$n$}
  \psfrag{(a,b)_2}{$0$}
  \psfrag{(a,b)_3}{$0$}
  \psfrag{(a,b)_4}{$n/2-2$}
  \psfrag{t0}{-2}
  \psfrag{t1}{0}
  \psfrag{t2}{0}
  \psfrag{t3}{0}
  \includegraphics{pic/newToken.eps}
} \qquad \subfigure[One underload token detours to the lower path. 
\label{SubFig:B2}]{
  \psfrag{(a,b)_0}{$n$}
  \psfrag{(a,b)_1}{$n$}
  \psfrag{(a,b)_2}{$1$}
  \psfrag{(a,b)_3}{$0$}
  \psfrag{(a,b)_4}{$n/2-2$}
  \psfrag{t0}{-1}
  \psfrag{t1}{0}
  \psfrag{t2}{-1}
  \psfrag{t3}{0}
  \includegraphics{pic/newToken.eps}
} \\
\subfigure[It continues on the lower path \label{SubFig:C2}]{
  \psfrag{(a,b)_0}{$n$}
  \psfrag{(a,b)_1}{$n$}
  \psfrag{(a,b)_2}{$1$}
  \psfrag{(a,b)_3}{$1$}
  \psfrag{(a,b)_4}{$n/2-2$}
  \psfrag{t0}{-1}
  \psfrag{t1}{0}
  \psfrag{t2}{0}
  \psfrag{t3}{-1}
  \includegraphics{pic/newToken.eps}
} \hspace{0.5cm} \subfigure[\ldots and moves to the next gadget. 
\label{SubFig:D2}]{
  \psfrag{(a,b)_0}{$n$}
  \psfrag{(a,b)_1}{$n$}
  \psfrag{(a,b)_2}{$1$}
  \psfrag{(a,b)_3}{$1$}
  \psfrag{(a,b)_4}{$n/2-2$}
  \psfrag{t0}{-1}
  \psfrag{t1}{0}
  \psfrag{t2}{0}
  \psfrag{t3}{0}
  \includegraphics{pic/newToken.eps}
} \\
\subfigure[The second overload token moves. \label{SubFig:E2}]{
  \psfrag{(a,b)_0}{$n-1$}
  \psfrag{(a,b)_1}{$n-1$}
  \psfrag{(a,b)_2}{$1$}
  \psfrag{(a,b)_3}{$1$}
  \psfrag{(a,b)_4}{$n/2-1$}
  \psfrag{t0}{0}
  \psfrag{t1}{0}
  \psfrag{t2}{0}
  \psfrag{t3}{-1}
  \includegraphics{pic/newToken.eps}
} \hspace{0.5cm} \subfigure[New tokens are generated. \label{SubFig:F2}]{
  \psfrag{(a,b)_0}{$n$}
  \psfrag{(a,b)_1}{$n-1$}
  \psfrag{(a,b)_2}{$0$}
  \psfrag{(a,b)_3}{$1$}
  \psfrag{(a,b)_4}{$n/2-1$}
  \psfrag{t0}{-1}
  \psfrag{t1}{0}
  \psfrag{t2}{+1}
  \psfrag{t3}{0}
  \includegraphics{pic/newToken.eps}
}
\caption{The number of tokens increases along the lower path.}
\label{Fig:NewTokensUnderload}
\end{center}\end{figure}

Next we describe how the number of overload and underload tokens can
increase. This can happen if there are either at least two overload or at
least two underload tokens on $r_{i,0}$. In the following, we discuss the
first case in detail. The second case in only depicted in
Figure~\ref{Fig:NewTokensUnderload}.

\begin{figure}[ht] \begin{center}
\subfigure[Initial configuration. \label{SubFig:A1}]{
  \psfrag{(a,b)_0}{$n$}
  \psfrag{(a,b)_1}{$n$}
  \psfrag{(a,b)_2}{$0$}
  \psfrag{(a,b)_3}{$0$}
  \psfrag{(a,b)_4}{$n/2$}
  \psfrag{t0}{+2}
  \psfrag{t1}{0}
  \psfrag{t2}{0}
  \psfrag{t3}{0}
  \includegraphics{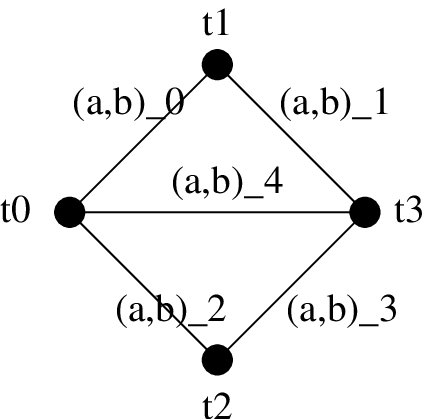}
} \qquad \subfigure[One overload token detours to the upper path. 
\label{SubFig:B1}]{
  \psfrag{(a,b)_0}{$n-1$}
  \psfrag{(a,b)_1}{$n$}
  \psfrag{(a,b)_2}{$0$}
  \psfrag{(a,b)_3}{$0$}
  \psfrag{(a,b)_4}{$n/2$}
  \psfrag{t0}{+1}
  \psfrag{t1}{+1}
  \psfrag{t2}{0}
  \psfrag{t3}{0}
  \includegraphics{pic/newToken.eps}
} \\
\subfigure[It continues on the upper path \label{SubFig:C1}]{
  \psfrag{(a,b)_0}{$n-1$}
  \psfrag{(a,b)_1}{$n-1$}
  \psfrag{(a,b)_2}{$0$}
  \psfrag{(a,b)_3}{$0$}
  \psfrag{(a,b)_4}{$n/2$}
  \psfrag{t0}{+1}
  \psfrag{t1}{0}
  \psfrag{t2}{0}
  \psfrag{t3}{+1}
  \includegraphics{pic/newToken.eps}
} \qquad \subfigure[\ldots and moves to the next gadget. \label{SubFig:D1}]{
  \psfrag{(a,b)_0}{$n-1$}
  \psfrag{(a,b)_1}{$n-1$}
  \psfrag{(a,b)_2}{$0$}
  \psfrag{(a,b)_3}{$0$}
  \psfrag{(a,b)_4}{$n/2$}
  \psfrag{t0}{+1}
  \psfrag{t1}{0}
  \psfrag{t2}{0}
  \psfrag{t3}{0}
  \includegraphics{pic/newToken.eps}
} \\
\subfigure[The second overload token moves. \label{SubFig:E1}]{
  \psfrag{(a,b)_0}{$n-1$}
  \psfrag{(a,b)_1}{$n-1$}
  \psfrag{(a,b)_2}{$0$}
  \psfrag{(a,b)_3}{$0$}
  \psfrag{(a,b)_4}{$n/2-1$}
  \psfrag{t0}{0}
  \psfrag{t1}{0}
  \psfrag{t2}{0}
  \psfrag{t3}{+1}
  \includegraphics{pic/newToken.eps}
} \qquad \subfigure[New tokens are generated. \label{SubFig:F1}]{
  \psfrag{(a,b)_0}{$n$}
  \psfrag{(a,b)_1}{$n-1$}
  \psfrag{(a,b)_2}{$0$}
  \psfrag{(a,b)_3}{$0$}
  \psfrag{(a,b)_4}{$n/2-1$}
  \psfrag{t0}{+1}
  \psfrag{t1}{-1}
  \psfrag{t2}{0}
  \psfrag{t3}{}
  \includegraphics{pic/newToken.eps}
}
\caption{The number of tokens increases along the upper path.} \label{Fig:NewTokens1}
\end{center}\end{figure}

Consider a single gadget $G_i$ as depicted in Figure~\ref{SubFig:A1}. Numbers 
attached to resources correspond to the number of tokens lying on them. 
Positive numbers indicate that overload tokens are present, negative numbers 
indicate that underload tokens are present. Numbers $a$ attached to edges 
indicate that $a$ players represented by that edge play their 0-strategy, 
whereas $n-a$ players play their 1-strategy.

\begin{description}
  \item[Configuration~\ref{SubFig:A1}:] Initially, there are two overload 
  tokens on $r_{i,0}$. In this case, all type 0 and all type 4 players have an 
  incentive to change to their 1-strategies. All other players are satisfied. 
  With probability $2/3$, given that a player from $G_i$ is selected, a type 0 
  player is selected and the configuration~\ref{SubFig:B1} is reached,
  in which there is one overload token on $r_{i,0}$ and one on $r_{i,1}$.
  \item[Configuration~\ref{SubFig:B1}:] All type 1 and all type 4 players have 
  an incentive to change to their 1-strategy. With probability $2/3$ 
  configuration~\ref{SubFig:C1} is reached in which there is one overload token 
  on $r_{i,0}$ and one on $r_{i,3}$.
  \item[Configuration~\ref{SubFig:C1}:] Still all type 4 players have an 
  incentive to change to their 1-strategy. However, we assume that the overload 
  token which currently lies on $r_{i,0}$ moves on due to a best response of a 
  player in gadget $G_{i+1}$. In this case, configuration~\ref{SubFig:D1} is 
  reached in which there is still one overload token on $r_{i,0}$. 
  Additionally, one overload token is in gadget $G_{i+1}$.
  \item[Configuration~\ref{SubFig:D1}:] Again, all type 4 players have an 
  incentive to change to their 1-strategy. Now one of these players is selected 
  and configuration~\ref{SubFig:E1} is reached in which there is one overload 
  token on $r_{i,4}$.
  \item[Configuration~\ref{SubFig:E1}:] In this configuration, the overload 
  token on $r_{i,4}$ can move to the next gadget. Observe that this event is 
  much more likely than the next one, in which the only type 0 player playing 
  its 1-strategy switches back to its 0-strategy. All other players are 
  satisfied. If both events take place configuration~\ref{SubFig:F1} is 
  reached. Note, that in this case additional tokens are generated.
  There is a new underload token on $r_{i,1}$ and a new overload token on $r_{i,0}$.
  \item[Configuration~\ref{SubFig:F1}:] Finally, all $n-1$ type 4 players 
  playing their 0-strategy have an incentive to change to their 1-strategy. 
  Additionally, the only type 1 player playing its 1-strategy wants to change 
  back to its its 0-strategy.
\end{description}

\subsection{Simulations}

\begin{figure}[ht] \begin{center}
  \includegraphics[scale=0.8]{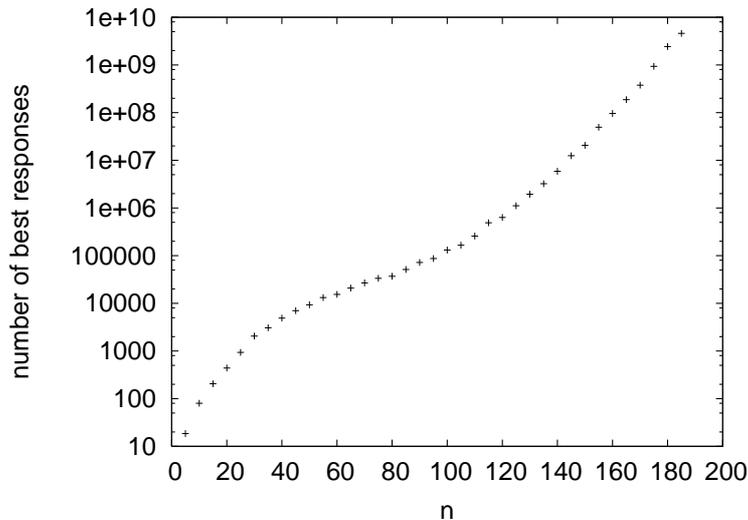}
  \caption{Average number of best responses}
  \label{Fig:plottedRunningTime}
\end{center}\end{figure}

We simulated the random best response dynamics in games $\Gamma_n$ and
obtained the results shown in Figure~\ref{Fig:plottedRunningTime}. On the
$x$-axis we plotted the parameter $n$, on the $y$-axis the average number
of best responses until the random best response dynamics terminated.
Observe that the $y$-axis is plotted in $\log$-scale. For every $n \in
\{5,10,\ldots,180,185\}$ we started the random best response dynamics
from the following initial configuration: all type 0 and all type 1
players play their 0-strategies; all type 2 and all type 3 players play
their 1-strategies. Additionally, $n/2$ type 4 players in the gadgets
$G_0,\ldots,G_{n/2-1}$ and $n/2+2$ type 4 players in the gadgets
$G_{n/2},\ldots,G_{n-1}$ play their 1-strategy. All other type 4 players
play their 0-strategy. This initial configuration corresponds to placing
two overload tokens on $r_{0,0}$ and two underload tokens on $r_{n/2,0}$.
For $n \le 160$ we took the average over 400 runs, and for larger $n$ we
took the average over 100 runs.

Unfortunately, it does not seem feasible to simulate the best
response dynamics for much larger values of $n$. We believe, however, that the
results in Figure~\ref{Fig:plottedRunningTime} are a clear indication
for a super-polynomial, maybe even exponential, convergence time.

  
\section{Acknowledgment}
The authors wish to thank Berthold V\"ocking for valuable discussions.

  \bibliographystyle{plain}
  \bibliography{bibSAGA}

\end{document}